\documentclass[journal=jacsat,manuscript=article]{achemso}
\pdfoutput=1 
\usepackage[version=3]{mhchem} 

\author{Linzhi Yu}
\affiliation{Department of Physics, Tampere University, 33720, Tampere, Finland}
\author{Sergei Shevtsov}
\affiliation{Optoelectronics Research Centre, University of Southampton, Southampton SO17 1BJ, UK}
\author{Haobijam Johnson Singh}
\affiliation{Department of Physics, Tampere University, 33720, Tampere, Finland}
\author{Peter G. Kazansky}
\affiliation{Optoelectronics Research Centre, University of Southampton, Southampton SO17 1BJ, UK}
\author{Humeyra Caglayan}
\affiliation{Department of Physics, Tampere University, 33720, Tampere, Finland}
\alsoaffiliation[Second University]{Department of Electrical Engineering and Eindhoven Hendrik Casimir Institute, Eindhoven University of Technology, Eindhoven 5600 MB, The Netherlands}
\email{h.caglayan@tue.nl}
\title[An \textsf{achemso} demo]
  {Multifunctional meta-optic azimuthal shear interferometer}

\abbreviations{IR,NMR,UV}
\keywords{American Chemical Society, \LaTeX}

\begin{document}

\begin{abstract}
  Azimuthal shear interferometry is a versatile tool for analyzing wavefront asymmetries. However, conventional systems are bulky, alignment-sensitive, and prone to nonuniform shear. We present a broadband, compact, and robust meta-optics-based azimuthal shear interferometer in a common-path configuration, reducing the system size to the millimeter scale. Unlike conventional designs, the meta-optic azimuthal shear interferometer utilizes the localized wavefront modulation capabilities of meta-optics to achieve uniform azimuthal shear displacement independent of radial position, significantly enhancing accuracy and stability. Our approach eliminates the need for bulky optical components and precise multi-path alignment, making it more resilient to environmental disturbances. Its multifunctionality is demonstrated through applications in all-optical edge detection, differential interference contrast microscopy, and aberrated wavefront sensing. These results underscore its potential for real-time analog image processing, advanced optical imaging, and optical testing.
\end{abstract}

\section{Introduction}
Shear interferometry is a versatile technique for analyzing the amplitude or phase of a light wavefront by interfering the wavefront with its displaced duplicate. This approach has been widely applied in areas such as optical component testing, phase imaging, and fluid mechanics. Based on the displacement strategy, it can be classified into lateral, radial, azimuthal, and reversal shear~\cite{OpticalShopTesting,rimmer1975evaluation,murty1966rotational}. Among these, azimuthal shear interferometry is particularly notable for its high sensitivity to asymmetrical wavefront features. Conventional azimuthal shear interferometry systems duplicate wavefronts using a beam splitter, rotate the duplicates along the optical axis using Dove prisms, and recombine them via a second beam splitter. The resulting interference pattern reveals asymmetries in the original wavefront. While effective, these systems face significant challenges. Their bulky designs, dependence on precise alignment, and sensitivity to environmental instabilities arising from their separate path configurations—limit their practicality. Furthermore, nonuniform interference fields result from shear amounts being proportional to radial positions (see Supporting Information(S2)), undermining accuracy and robustness in practical applications~\cite{murty1966rotational,bryngdahl1974shearing,OpticalShopTesting}. Addressing these limitations requires a more compact, stable, and uniform solution.

Meta-optics, planar structures composed of subwavelength-scale resonators (meta-atoms), enable precise and independent modulation of electromagnetic fields~\cite{yu2011light,kuznetsov2024roadmap}. These advanced components have revolutionized optical design by facilitating the miniaturization and enhancement of traditional optical elements, such as lenses~\cite{pan2022dielectric}, waveplates~\cite{deng2022recent}, and polarizers~\cite{wang2023metasurface}. Moreover, meta-optics provide unprecedented multifunctionality and demonstrate strong performance in areas such as all-optical signal processing, optical imaging, and wavefront sensing. Recent research has demonstrated the remarkable potential of metasurfaces for various applications. For instance, metasurfaces based on nonlocality~\cite{guo2018photonic,cordaro2019high,zhou2020flat,cotrufo2023polarization,cotrufo2023dispersion,de2023analog,liu2024edge,esfahani2024tailoring,zhou2024laplace,cotrufo2024reconfigurable}, spiral phase filtering~\cite{huo2020photonic,kim2022spiral,wang2023metalens,zhang2023dielectric}, and differential interference contrast (DIC)~\cite{zhou2019optical,zhou2020metasurface,zhou2021two,wang2023single,zhou2025meta}, among others~\cite{swartz2024broadband,tanriover2023metasurface,zhou2022nonlinear}, have enabled real-time, all-optical edge detection, addressing challenges such as low processing speeds and high power consumption in machine vision systems. Moreover, principles such as nonlocality~\cite{ji2022quantitative,zhou2020flat}, DIC~\cite{kwon2020single,wang2023single,li2024single,wu2023single}, spiral phase filtering~\cite{huo2020photonic,kim2022spiral,wang2023metalens,zhang2023dielectric,deng2024dielectric}, and transport intensity equations~\cite{engay2021polarization,zhou2024eagle} have been effectively leveraged by meta-optics for phase object imaging. Despite their promise, these systems are often constrained by inherent limitations, including wavelength dependence~\cite{guo2018photonic,cordaro2019high,zhou2020flat,cotrufo2023polarization,cotrufo2023dispersion,de2023analog,liu2024edge,esfahani2024tailoring,zhou2024laplace,cotrufo2024reconfigurable,ji2022quantitative}, insensitivity to asymmetrical features~\cite{zhou2019optical,zhou2020metasurface,kwon2020single,zhou2021two,wang2023single,wu2023single,li2024single}, and nonuniform background artifacts~\cite{huo2020photonic,kim2022spiral,wang2023metalens,zhang2023dielectric,deng2024dielectric}.

This work introduces a birefringent meta-optics-based azimuthal shear interferometer (meta-ASI), which achieves azimuthal shear interference in a compact, millimeter-scale common-path configuration with broadband operation. Unlike conventional systems that suffer from nonuniform shear, alignment sensitivity, and bulkiness, as well as existing meta-optics approaches that are often wavelength-dependent or insensitive to asymmetrical wavefronts, the meta-ASI leverages precise, localized modulation to achieve uniform azimuthal shear displacement independent of radial position. The versatility of the system is demonstrated experimentally through its applications in all-optical image edge detection, DIC microscopy imaging, and azimuthal shear interference of random aberrated wavefronts. These results establish the meta-ASI as a transformative platform to enhance edge contrast in images without post-processing, e.g., in biological specimens, and analyze wavefront distortions in real-time for applications in adaptive optics and high-resolution wavefront sensing.

\section{Results and discussion}

\begin{figure}[h!]
  \centering\includegraphics[width=16cm]{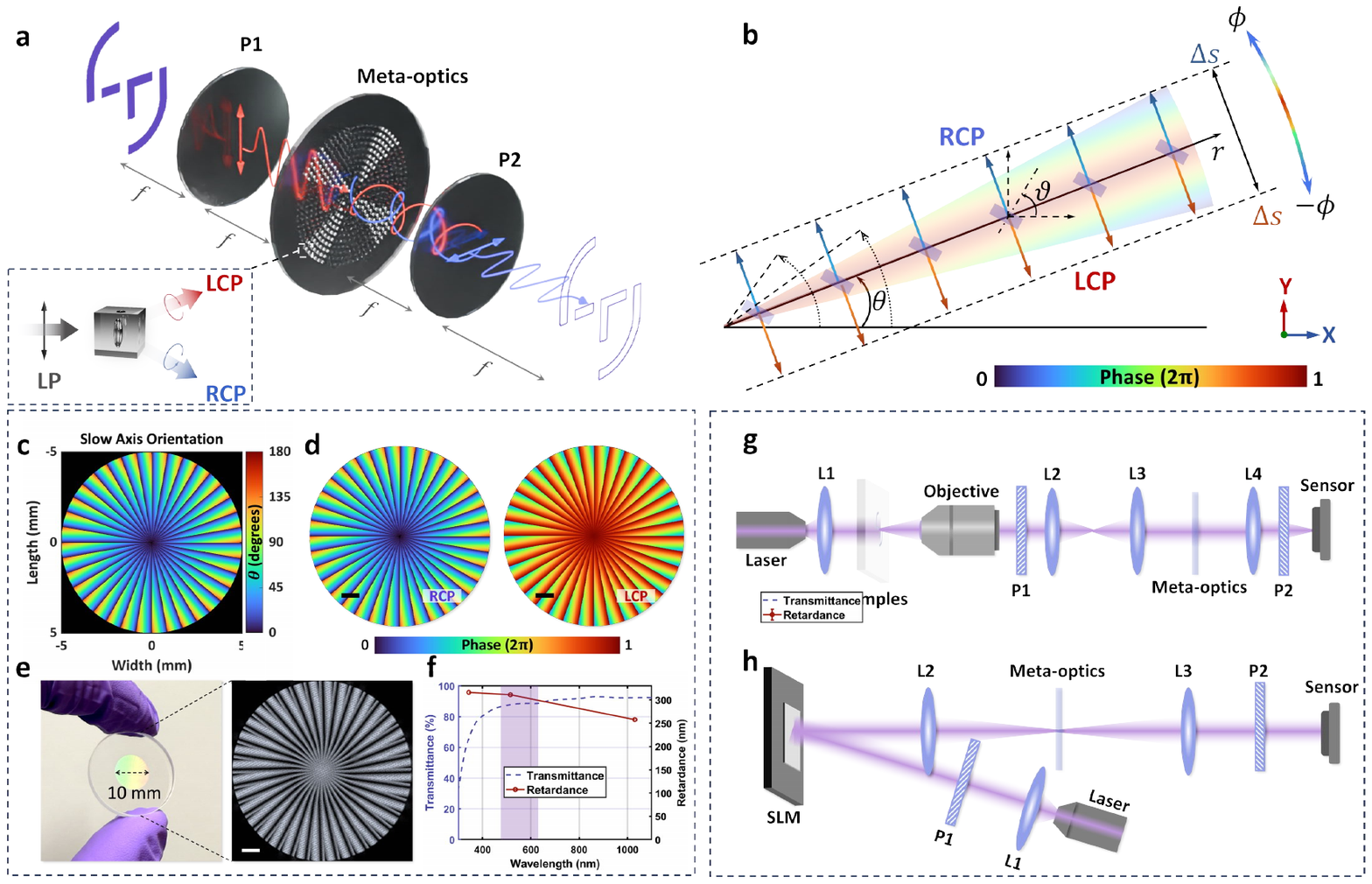}
  \caption{Design and experimental setup of the meta-ASI. (a) Conceptual illustration of the meta-ASI. (b) Design principle of the meta-optics. (c) Short-axis orientation distribution of nano half-waveplates on the designed meta-optic. (d) Phase map distributions of the meta-optic for RCP and LCP. (e) Fabricated meta-optic and its birefringent image. (f) Transmittance and phase retardance spectra of the meta-optic. The shaded region represents the spectral range used in the experiments presented in this work. (g) Experimental setup for all-optical image edge detection and DIC microscopy. (h) Experimental setup for azimuthal shear interference of wavefronts. L indicates a lens, and P indicates a polarizer. Scale bars in (d,e) represent 1 mm.}
  \label{fgr:fig1}
\end{figure}

Meta-ASI consists of a planar meta-optic positioned at the Fourier plane of an optical system, as shown in Figure~\ref{fgr:fig1}a. The meta-optic comprises birefringent meta-atoms that function as half-wave plates, independently modulating the light wavefront. Each meta-atom decomposes an incident linearly polarized (LP) wavefront into right-handed circularly polarized (RCP) and left-handed circularly polarized (LCP) components, imposing opposite phase delays \(+2\vartheta\) and \(-2\vartheta\), respectively, where \(\vartheta\) is the short-axis orientation angle of the birefringent meta-atom (Figure~\ref{fgr:fig1}b; detailed in Supporting Information(S1)). This phenomenon, known as the photonic spin Hall effect~\cite{yin2013photonic,liu2022photonic}, is illustrated in the inset of Figure~\ref{fgr:fig1}a. Leveraging this property, the meta-atoms act as beam splitters in the meta-ASI, duplicating the wavefronts into RCP and LCP components with opposite azimuthal shear displacements. The azimuthal shear displacements \(\pm \Delta s\) are achieved on the image plane of the Fourier lens by deflecting the wavefronts along the azimuthal direction on the Fourier plane, where \(\pm \Delta s = \pm \lambda f \frac{1}{r} \frac{\partial \phi}{\partial \theta}\). Here, \(\lambda\) is the light wavelength, \(f\) is the focal length of the Fourier lens, and \(\pm \phi(r,\theta)\) represents the phase delay distribution applied to RCP and LCP components, respectively~\cite{goodman2017introduction}. Since achieving a constant azimuthal displacement across different radii \(r\) is impractical for planar optics (as illustrated by the dotted arcs in Figure~\ref{fgr:fig1}b), a tangent displacement is introduced instead, using a phase delay distribution of \(\pm \phi(r,\theta) = \pm r \theta C\). This ensures a constant azimuthal displacement \(\pm \Delta s = \pm \lambda f C\), where \(C\) controls the shear amount. However, this phase distribution also introduces a radial displacement \(\pm \Delta r = \pm \lambda f \frac{\partial \phi}{\partial r} = \pm \lambda f \theta C\), which causes a spiral error proportional to \(\theta\). To mitigate this error, the meta-optic is divided into \(N\) azimuthal sections, with \(\theta\) ranging from \(0\) to \(\frac{2\pi}{N}\) in each section. Increasing  \(N\) reduces the error asymptotically, as detailed in Supporting Information(S3). When an LP wavefront \(E(r,\theta)\) is incident, the RCP and LCP components are modulated by the phase distributions \(\phi(r,\theta)\) and \(-\phi(r,\theta)\), as determined by the short-axis orientation of meta-atoms \(\vartheta(r,\theta) = \frac{\phi(r,\theta)}{2}\). This results in a shear field with uniform azimuthal displacement, leading to the intensity distribution:
\begin{equation}
I(r,\theta) \propto \left|E\left(r,\theta+\frac{\Delta s}{r}\right) - E\left(r,\theta-\frac{\Delta s}{r}\right)\right|^2,
\label{eq:eq1}
\end{equation}
For small \(\Delta s\), applying Taylor expansion yields:
\begin{equation}
I(r,\theta) \propto 2 \Delta s \left|\frac{1}{r} \frac{\partial E}{\partial \theta} \right|^2,
\label{eq:eq2}
\end{equation}
where \(\Delta s\) is constant, ensuring uniform azimuthal shear across the wavefront. This uniform shear field intensity directly corresponds to the azimuthal gradient of the complex incident wavefront. This approach ensures consistent shear performance, addressing challenges inherent to conventional designs.

To validate the concept, a meta-ASI with a diameter of 10 mm was designed. The shear power was specified as \(C = 5.9 \times 10^3 \, \mathrm{m}^{-1}\), and the number of azimuthal sections \(N\) was set to \(36\) to balance radial error and the resolution of the meta-optic. The orientation distribution of the slow axes of the meta-atom is illustrated in Figure~\ref{fgr:fig1}c. Based on the photonic spin Hall effect, the phase responses for the RCP and LCP components of this meta-ASI are shown in Figure~\ref{fgr:fig1}d. The meta-ASI was fabricated using laser writing technology on a fused silica substrate~\cite{shimotsuma2003self,sakakura2020ultralow,Lei:23,kazansky2024}. This fabrication process ensures high transmittance across a broad wavelength range. The fabricated meta-ASI and its birefringent image are shown in Figure~\ref{fgr:fig1}e. Figure~\ref{fgr:fig1}f presents the transmittance and phase retardance spectra from 300 nm to 1100 nm, demonstrating over 80\% transmittance in the visible range. Due to its geometric phase modulation strategy, the meta-ASI avoids dispersion-induced phase errors while only reducing efficiency, making it highly suitable for broadband applications. Further details can be found in the Methods and Supporting Information(S6).

\begin{figure}[h!]
  \centering\includegraphics[width=13.92cm]{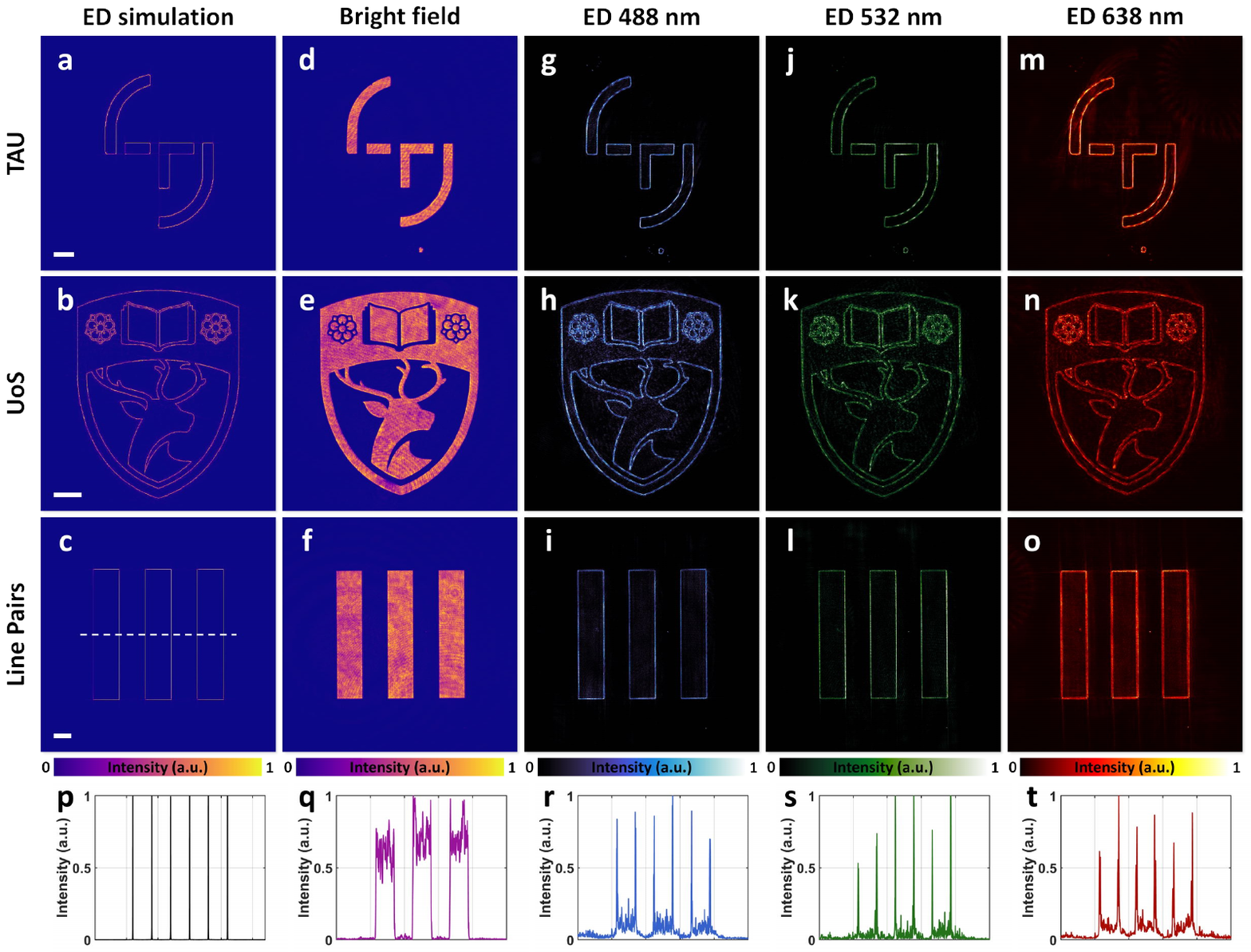}
  \caption{Meta-ASI for edge detection (ED) of amplitude images. The first, second, and third rows correspond to images with TAU, UoS, and resolution line pair patterns, respectively. (a–c) Simulated ED results. (d–f) Experimental results under bright-field illumination. (g–i) Experimental ED results under blue light illumination (488 nm). (j–l) Experimental ED results under green light illumination (532 nm). (m–o) Experimental ED results under red light illumination (638 nm). (p–t) Cross-sectional intensity profiles corresponding to (c), (f), (i), (l), and (o). Scale bars: 200 \textmu m.}
  \label{fgr:fig2}
\end{figure}

We first characterized the meta-ASI's capability for all-optical image edge detection using both amplitude and phase images. The test images, shown in Figures S1(a-c), Supporting Information, include the logos of Tampere University and the University of Southampton, as well as a resolution line pair pattern. We first evaluated edge detection for amplitude images, where the patterned regions are transmissive and other areas are opaque. The numerically simulated edge-detected results for the ideal meta-ASI design are presented in Figures~\ref{fgr:fig2}a-c. The experimental setup, shown in Figure~\ref{fgr:fig1}g, was used for testing. Without the meta-ASI in the system, the sensor captured images (Figures~\ref{fgr:fig2}d-f) that exhibited slight nonuniform brightness on the patterns due to imperfect illumination and laser noise. After placing the meta-ASI at the 4f plane, the sensor recorded azimuthally sheared light fields. To evaluate the broadband performance of the meta-ASI, each image sample was illuminated at three wavelengths: 488 nm (blue), 532 nm (green), and 638 nm (red). The experimentally captured edge-detected images are shown in Figures~\ref{fgr:fig2}g-o, demonstrating effective edge detection across all three wavelengths. For quantitative analysis, Figures~\ref{fgr:fig2}p-t presents normalized cross-sectional intensity profiles of the results for the resolution line pair pattern shown in Figures~\ref{fgr:fig2}c,f,i,l,o. These profiles reveal sharp edge extraction with minimal background intensity noise.

\begin{figure}[h!]
  \centering\includegraphics[width=13.92cm]{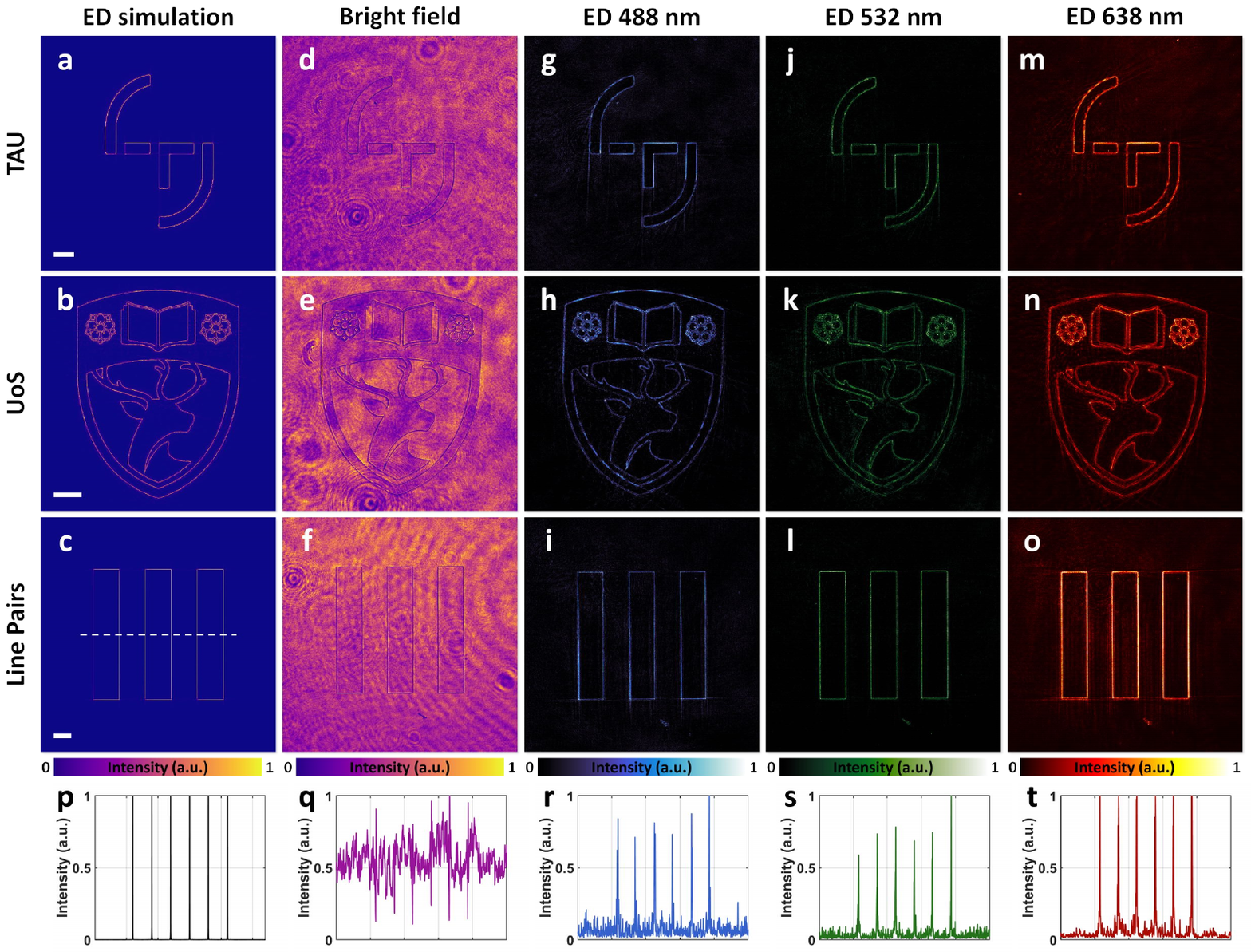}
  \caption{Meta-ASI for edge detection (ED) of phase images. The first, second, and third rows correspond to images with TAU, UoS, and resolution line pair patterns, respectively. (a–c) Simulated ED results. (d–f) Experimental results under bright-field illumination. (g–i) Experimental ED results under blue light illumination (488 nm). (j–l) Experimental ED results under green light illumination (532 nm). (m–o) Experimental ED results under red light illumination (638 nm). (p–t) Cross-sectional intensity profiles corresponding to (c), (f), (i), (l), and (o). Scale bars: 200 \textmu m.}
  \label{fgr:fig3}
\end{figure}

Since the meta-ASI manipulates the complex light field, it is also effective for edge detection to phase images, where patterned regions exhibit varying optical phase delays compared to the background. Theoretical edge-detected images from numerical simulation are shown in Figures~\ref{fgr:fig3}a-c. Experiments were conducted using the same setup Figure~\ref{fgr:fig1}g. Without the meta-ASI, the sensor captured bright-field images Figures~\ref{fgr:fig3}d-f, where phase patterns were barely visible. With the meta-ASI aligned, the sensor recorded azimuthally sheared light fields under illumination at 488 nm, 532 nm, and 638 nm. The resulting images, shown in Figures~\ref{fgr:fig3}g-o, clearly reveal the previously invisible edges of the phase patterns across all wavelengths. Normalized cross-sectional intensity profiles of Figures~\ref{fgr:fig3}c,f,i,l,o are shown in Figures~\ref{fgr:fig3}p-t for quantitative analysis. These profiles demonstrate sharp edge enhancement achieved with the meta-ASI. The broadband, real-time image edge detection capability of the meta-ASI holds significant potential for high-speed, low-energy image processing applications, such as intelligent manufacturing and automated driving~\cite{zangeneh2021analogue}.

\begin{figure}[h!]
  \centering\includegraphics[width=11.16cm]{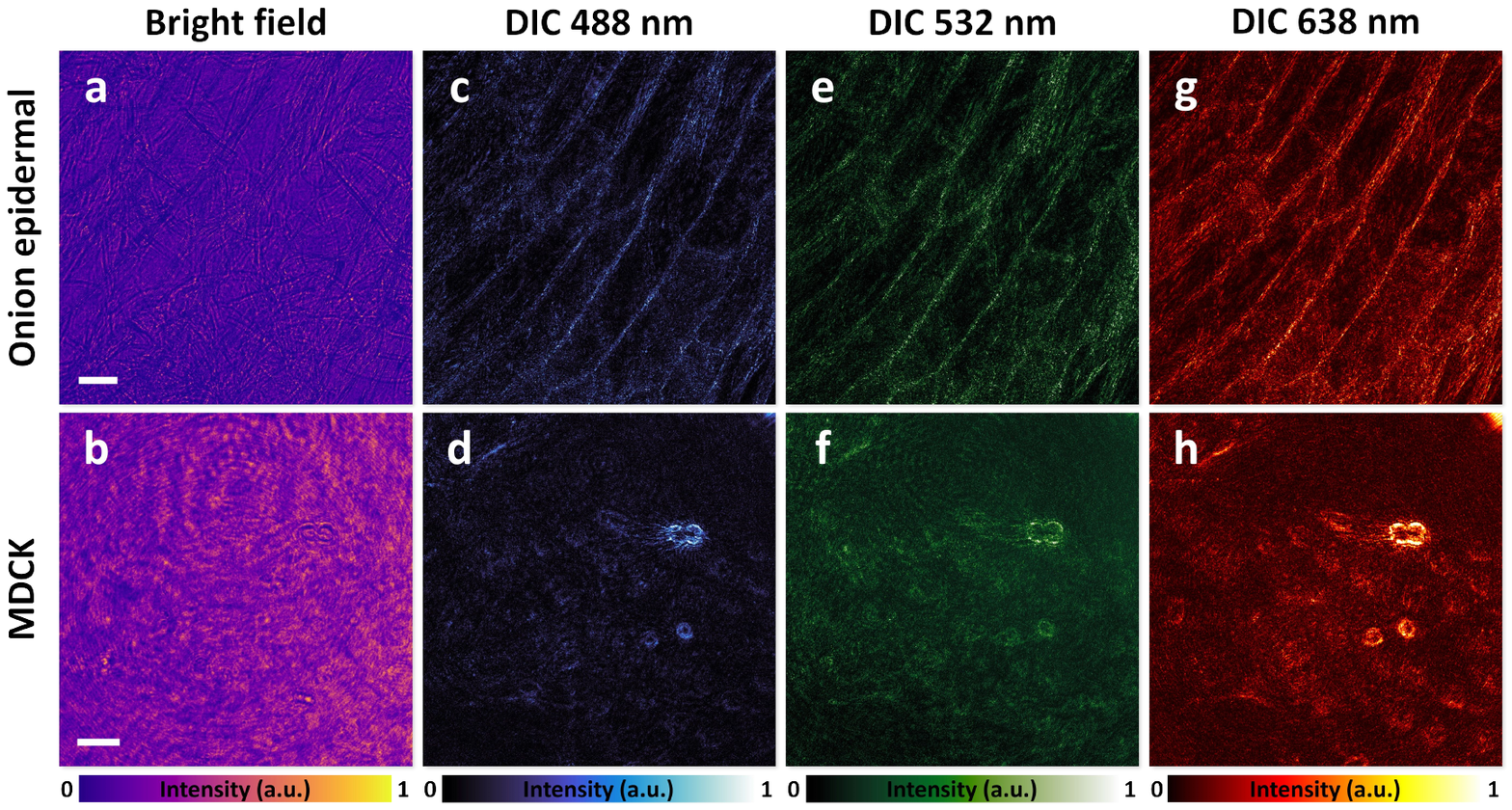}
  \caption{Meta-ASI for DIC microscopy. (a, b) Bright-field images of onion epidermal cells and MDCK cells, respectively. (c, e, g) DIC images of onion epidermal cells under blue (488 nm), green (532 nm), and red (638 nm) light illumination. (d, f, h) DIC images of MDCK cells under blue, green, and red light illumination, respectively. Scale bars for the onion epidermal cell images: 200 \textmu m. Scale bars for the MDCK cell images: 50 \textmu m.}
  \label{fgr:fig4}
\end{figure}

In addition to its application in all-optical image edge detection, the meta-ASI can also serve as an ultra-compact DIC microscope~\cite{Preza:99,arnison2004linear}, an indispensable tool in biomedical science. To demonstrate this capability, we used the experimental setup shown in Figure~\ref{fgr:fig1}g, replacing the objective lenses with those of larger numerical apertures. Biological samples were observed, including onion epidermal cells and Madin-Darby canine kidney (MDCK) cells. Both types of cells are transparent but exhibit refractive index variations due to their organelles. Illumination light wavelengths of 488 nm (blue), 532 nm (green), and 638 nm (red) were used. Under bright-field illumination, the cells were barely visible, as shown in Figures~\ref{fgr:fig4}a,b. Introducing the meta-ASI significantly enhanced the contrast of the cells across all wavelengths, as shown in Figures~\ref{fgr:fig4}c-h. Compared to conventional DIC microscopy, which uses bulky birefringent prisms to duplicate and displace wavefronts along a single direction, the meta-ASI offers a more compact solution. Moreover, because the shear occurs along the azimuthal direction in polar coordinates rather than a single linear direction, the contrast enhancement is uniform. This design enables portable and scalable DIC microscopy, making it particularly valuable for point-of-care diagnostics and field-based biological research.

\begin{figure}[h!]
  \centering\includegraphics[width=11.26cm]{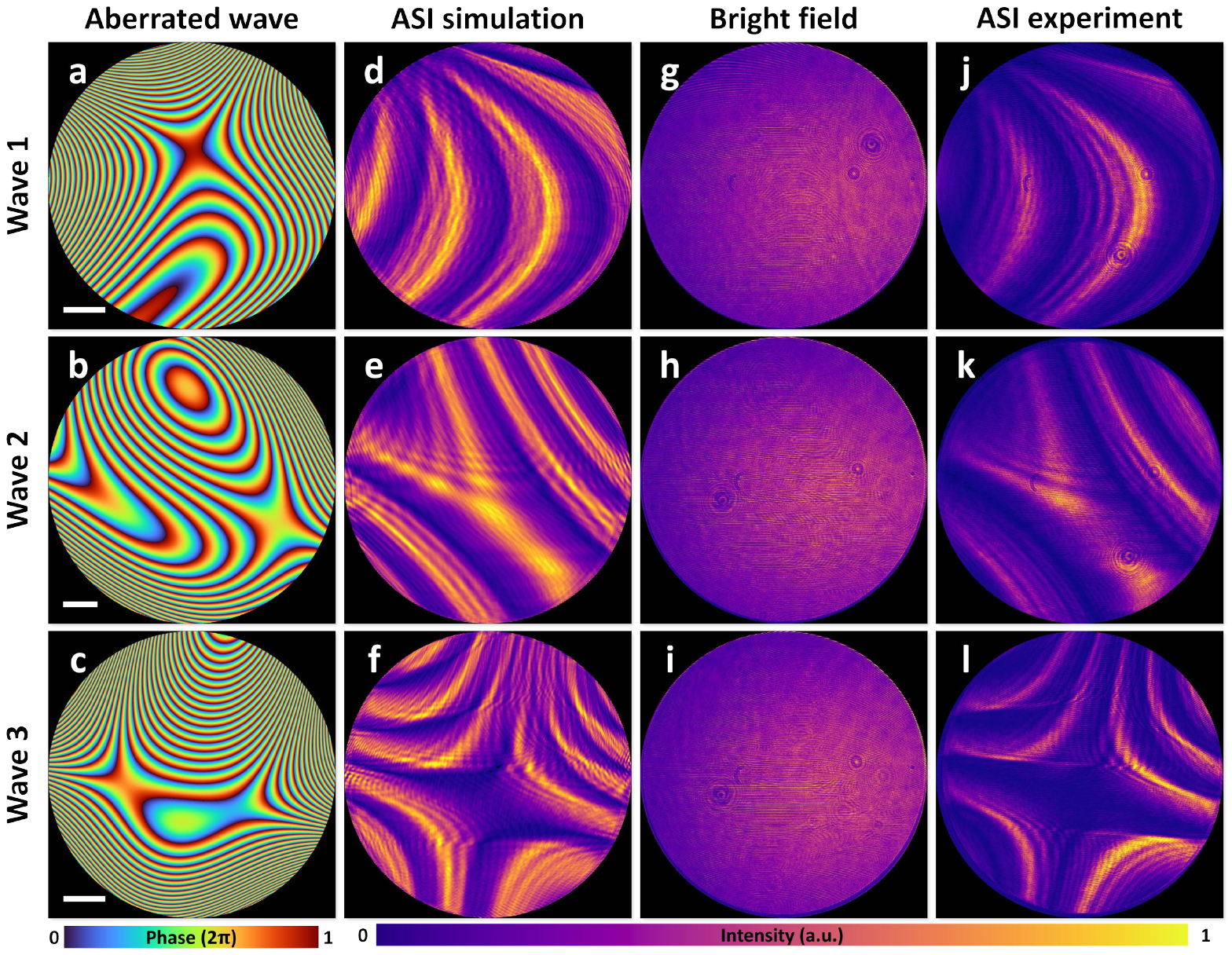}
  \caption{Meta-optic azimuthal shear interference (ASI) for aberrated wavefronts. (a, e, i) Phase maps of aberrated wavefronts 1, 2, and 3. (b, f, j) Simulated ASI intensity distributions. (c, g, k) Experimental intensity distributions without ASI. (d, h, l) Experimental ASI intensity distributions. Scale bars: 1 mm.}
  \label{fgr:fig5}
\end{figure}

To further illustrate the versatility and power of the meta-ASI, we demonstrated its use for sensing aberrated wavefronts through azimuthal shear interference. This technique converts the shape of the wavefront into a specific interference pattern determined by the shear amount. In this study, wavefronts were generated based on the 4th to 10th terms of Zernike polynomials, which represent common low-order aberrations~\cite{wang1980wave,zhao2007orthonormal,OpticalShopTesting}. Random coefficients were assigned to each term, and the corresponding phase maps are shown in Figures~\ref{fgr:fig5}a-c (details are provided in Supporting Information(S4)). The theoretical intensity distributions of the azimuthal shear interference fields were obtained through simulation and are presented in Figures~\ref{fgr:fig5}d-f. Experimental validation was conducted using the setup depicted in Figure~\ref{fgr:fig1}h. Without the meta-ASI, the light field intensities captured by the sensor are shown in Figures~\ref{fgr:fig5}g-i, displaying no discernible patterns. When the meta-ASI was introduced, the captured intensity distributions, shown in Figures~\ref{fgr:fig5}j-l, revealed clear interference patterns. The simulated and experimental results show excellent agreement, confirming the meta-ASI's effectiveness in sensing aberrated wavefronts. This capability has potential applications in various optical measurement scenarios, particularly for microsystems~\cite{osten2018optical} and microfluidic~\cite{yang2018micro}, where conventional setups, such as interferometers, are too bulky and complex reference wavefronts are challenging to generate. Additionally, the broadband property of the meta-ASI makes it robust against dispersion effects, further enhancing its practicality for diverse optical applications.

\section{Conclusion}

In summary, we have introduced a broadband, ultra-compact azimuthal shear interferometer enabled by meta-optics. By leveraging the ability of meta-optics to split and independently displace the LCP and RCP components, the system achieves a uniform azimuthally sheared field. This design not only miniaturizes conventional systems but also addresses the limitations of nonuniform shear through the precise, localized modulation capabilities of meta-optics. The multifunctionality of the meta-ASI has been demonstrated through diverse applications, including real-time all-optical image edge detection, DIC microscopy, and wavefront azimuthal shear interference across multiple wavelengths. Its compact and versatile design positions the meta-ASI as a transformative tool with significant potential for advancing real-time image processing, biomedical imaging, wavefront sensing, and optical testing.

\section{Methods}
\subsection{Numerical Simulation}
The theoretical intensity distributions for edge detection (Figures~\ref{fgr:fig2}a-c; Figures~\ref{fgr:fig3}a-c) and azimuthal shear interference (Figures~\ref{fgr:fig5}d-f) were simulated using Fourier optics~\cite{goodman2017introduction}. First, the amplitude transfer functions (ATFs) of the meta-ASI for RCP and LCP components were calculated. These ATFs were then multiplied by the spatial spectrum of the input light field to obtain the spatial spectrum of the output RCP and LCP fields. The final intensity distribution was derived by calculating the absolute square of the interfered fields. The meta-ASI design and all simulations were performed using MATLAB.

\subsection{Optical Characterization}
The meta-ASI's capabilities for all-optical edge detection and DIC microscopy were evaluated using the setup shown in Figure~\ref{fgr:fig1}g under illumination by laser beams at 488 nm, 532 nm, and 638 nm. The beams were expanded, collimated, and directed to illuminate the samples, which were imaged using an objective lens and a tube lens to form an intermediate image. This intermediate image was subsequently relayed through a 4f system onto an image sensor, with the meta-optic positioned at the Fourier plane. A polarizer placed before the intermediate image plane ensured that the incident wavefront was linearly polarized, while a second polarizer positioned between the 4f system and the sensor facilitated interference of the rotated RCP and LCP wavefronts. For azimuthal shear interference of aberrated wavefronts, the setup shown in Figure~\ref{fgr:fig1}h was used. A 532 nm polarized laser beam was directed onto a spatial light modulator and imaged through a 4f system onto an image sensor, with the meta-optic positioned at the Fourier plane. A polarizer was used to enable interference between the RCP and LCP components. Additional details are provided in Supporting Information(S5).

\subsection{Metasurface Fabrication}
The meta-optics was fabricated by laser direct writing inside a silica substrate using a mode-locked Yb:KGW ultrafast laser. Anisotropic nanoporous structures were imprinted with controlled slow-axis orientation via polarization modulation. A multilayer structure was implemented to achieve the required retardance, and the birefringent properties were characterized using an optical microscope with a birefringence measurement system. Further fabrication details are provided in Supporting Information(S6).

\subsection{Edge Detection Samples Fabrication}
Phase image samples were fabricated by spin-coating photoresist onto glass substrates, followed by maskless lithography, development, and post-baking to create transparent phase-modulating patterns. Amplitude image samples were prepared by spin-coating photoresist, developing the pattern, depositing a chromium layer, and performing a lift-off process to form transparent apertures. Further fabrication details are available in Supporting Information(S7).

\begin{acknowledgement}

L.Y. and H.C. acknowledge financial support from the European Union’s Horizon 2020 research and innovation programme under the Marie Skłodowska-Curie grant agreement No 956770. L.Y. thanks Rakesh Dhama, Anil Atalay Appak, Arttu Nieminen, and Jesse Pietilä for helpful discussions.

\end{acknowledgement}

\begin{suppinfo}

The theory of photonic spin Hall effect, performance analysis of meta-ASIs with varying angular shear amounts, performance analysis of meta-ASIs with varying azimuthal section numbers, aberrated wavefronts generation, experiment setups, meta-optics fabrication, and testing sample fabrication.

\end{suppinfo}

\bibliography{Reference}

\end{document}